# Modeling Complex Liquid Crystals Mixtures: From Polymer Dispersed Mesophase to Nematic Nanocolloids

Ezequiel R. Soule and Alejandro D. Rey

1. *Institute of Materials Science and Technology (INTEMA), University of Mar del Plata and National Research Council (CONICET), J. B. Justo 4302, 7600 Mar del Plata, Argentina*

2. *Department of Chemical Engineering, McGill University, Montreal, Quebec H3A 2B2, Canada*

## Abstract

Liquid crystals are synthetic and biological viscoelastic anisotropic soft matter materials that combine liquid fluidity with crystal anisotropy and find use in optical devices, sensor/actuators, lubrication, super-fibers. Frequently mesogens are mixed with colloidal and nanoparticles, other mesogens, isotropic solvents, thermoplastic polymers, cross-linkable monomers, among others. This comprehensive review present recent progress on meso and macro scale thermodynamic modelling, highlighting the (i) novelties in spinodal and binodal lines in the various phase diagrams, (ii) the growth laws under phase transitions and phase separation, (iii) the ubiquitous role of metastability and its manifestation in complex droplet interfaces, (iv) the various spinodal decompositions due to composition and order fluctuations, (v) the formation of novel material architectures such as colloidal crystals, (vi) the particle rich phase behaviour in liquid crystal nanocomposites, (vii) the use of topological defects to absorb and organize nanoparticles, and (viii) the ability of faceted nanoparticles to link into strings and organize into lattices. Emphasis is given to highlight dominant mechanisms and driving forces, and to link them to specific terms in the free energies of these complex mixtures. The novelties of incorporating mesophases into blends, solutions, dispersions and mixtures is revealed by using theory, modelling , computation, and visualization.

**1 - Introduction**

Liquid crystals (LCs) anisotropic viscoelastic soft matter materials displaying long range orientational and partial positional order, are responsive to electro-magnetic fields, substrates and interfaces, temperature and concentration gradients, and pH among many other fields [1-4]. They are found in many biological systems (DNA, cellular membrane, plant cell wall, etc.) [2;4;5] and form the basis of many functional and structural materials and devices, such as LCD displays, light valves, smart windows, sensor-actuators, artificial muscle, carbon super-fibers, among others [6-8]. There are many different types of liquid-crystalline phases (which are also known as mesophases); the simplest one is the nematic phase (N) that only displays orientational order.

In most technological applications and biological systems, the liquid crystalline material is not a pure substance but it is a mixture of two or more species, where at least one them is a liquid crystal. For example, in display applications, eutectic mixtures of different liquid crystals are used, in order to tune the temperature range where the N phase is stable [8] and reduce response time through optimization of viscoelastic properties. Another important example is cell division, where a model has been proposed that explains centriole formation in terms of a preferential segregation of a solute to the core of topological defects [9]. Mixtures of liquid crystals with other materials such as polymers [10-22] and colloidal particles [23-32] have been widely studied due to their interesting physics and potential for technological innovation. A mixture with a high mesogen concentration can remain homogeneous and behave essentially as a pure material (with modified properties), but the mixture can also phase-separate, leading to a heterogeneous material with complex morphologies and order parameter gradients that can significantly affect and improve its properties. Even the simplest binary mesogenic mixtures are characterized by conserved (concentration) and non-conserved (orientatioal order) order parameters whose couplings generate new thermodynamic instabilities and non-classical phase transition kinetics. Understanding the phase behavior and the dynamics of phase transitions and structure formation in mesogenic mixtures is thus a fundamental aspect of liquid crystal science and technology.

Theory and simulation of liquid crystal thermodynamics have contributed both to fundamental understanding and to practical applications. Molecular [33;34], mesoscopic [33], and macroscopic [3;35], LC models have been widely used and are now being integrated in multiscale simulation approaches [36]. In this review we focus on mesoscopic and macroscopic approaches describing the thermodynamics of phase equilibrium, phase transitions and structure formation in liquid crystal mixtures and composites, emphasizing the work of our group in nematic liquid crystals, in the last few years. In this type of models the state of order of the nematic phase is described by means of the quadrupolar tensor order parameter **Q**, defined as [1;3;4;35]:

$$\mathbf{Q} = S\left(\mathbf{nn} - \frac{\boldsymbol{\delta}}{3}\right) + \frac{P}{3}(\mathbf{ll} - \mathbf{mm}) \qquad (1)$$

where $S$ is the scalar uniaxial order parameter, $P$ is the biaxial order parameter, $\delta$ is the unit tensor and **n**, **l** and **m** are the eigenvectors of **Q**. The scalar uniaxial order parameter $S$ measures the degree of molecular alignment along the average orientation **n**. The biaxial order parameter measures the deviation of the molecular alignment distribution from axial symmetry, even uniaxial phases (where $P=0$ in equilibrium), can display biaxiality under the effect of external fields and surfaces and in the vicinity of topological defects.

The organization of this review is as follows. In section 2 we describe the calculation of phase diagrams, including phase coexistence (binodal) and stability limit (spinodal) lines. Section 2.1 deals with mixtures of polymers and LCs, section 2.2 with mixtures of two LCs and section 2.3 with dispersion of nanoparticles (NP) in LCs. Section 3 describes the dynamics of phase transition highlighting the effect of mixed conserved and non-conserved order parameters. In section 3.1 the dynamics of an interface in a phase transition, describing a nucleation and growth process, is analyzed, while in section 3.2 spinodal decomposition is discussed. Section 4 analyzes different aspects of structure formation; in section 4.1 the phase morphology in a phase separated system is presented, section 4.2 discusses the defect structure in a mixture and section 4.3 presents textures and defect configurations in particle-filled nematics. Finally, section 5 presents the conclusions.

**2 – Phase equilibrium**

The starting point for the calculation of phase diagrams is an expression for the free energy density, $f$, as a function of the relevant thermodynamic variables. As the free energy in an unperturbed bulk system does not depend on the direction of the liquid crystal orientation, the scalar order parameters suffice to describe the thermodynamic state of the system. In a uniaxial material the relevant thermodynamic variables are the nematic uniaxial order parameter $S$, the liquid crystal concentration $\phi$ and the temperature $T$. The equilibrium condition is that the chemical potentials of each component are the same in each phase. In addition, the free energy in each phase has to be minimal with respect to $S$. Traditionally, phase diagrams of mixtures involving a low molecular weight nematic LC and another substance are described by combining the Flory-Huggins theory of mixing [37-41] and the Maier-Saupe theory of nematic ordering [10;12;17;20;42;43], or some of their generalizations or modifications, although other approaches have been used too. Maier Saupe´s model describes the system in terms of an anisotropic energetic interaction, neglecting excluded volume effects, so it is suited for short LC molecules (low molar mass mesogens). High molecular weigth LCs including fibers and platelets have been modeled with Onsager´s model [44-46], which is based in excluded volume and it is strictly valid for a infinite aspect ratios. Combinations of both approaches (excluded volume plus energetic interactions) have been proposed too [10;17]. A different approach was taken by Flory, who derived a theory based on a lattice model [47-49]. A rich variety of phase diagrams have been described in the literature with different degrees of complexity, generic phase diagrams of the simplest mixture (LC + isotropic substance), are shown in fig. 1 and will be described in the following paragraphs and sections.

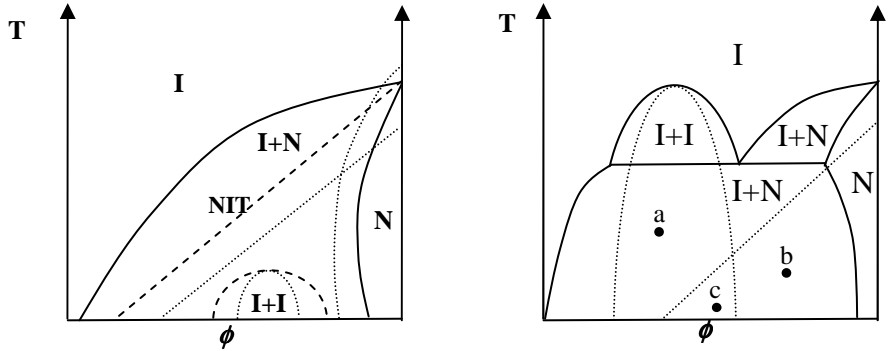

**Figure 1.** Schematic of phase diagrams of mixtures containing a nematic LC and a non-nematogenic species. I and N indicate isotropic and nematic phases. Full lines indicate binodals, dotted lines are spinodals. (a) with buried (metastable) I-I phase coexistence, dashed lines indicate the NIT line and the metastable I-I binodal (b) with I-I phase coexistence (The NIT line and the nematic spinodal are not shown for clarity reasons). Positions indicated as a, b, c show different quenches leading to different spinodal processes as discussed in section 3.b.

As shown in fig.1 Several regions can be defined in a phase diagram of mixtures involving LCs. Firstly, the $T$-$\phi$ diagram can be divided in a region where the isotropic phase is the equilibrium state of an homogeneous solution, and another region where the nematic phase is the equilibrium state. The limit between these two regions is called nematic-isotropic transition (NIT) line [10;17], and represents the first order transition of an homogeneous mixture, when phase separation is not allowed. This line represents an "extension" of the transition temperature of the pure LC, $T_{NI}$, to the mixture. When phase separation is allowed, the equilibrium phases at a given temperatures are given by binodal lines, and the region comprised inside a binodal correspond to phase coexistence. In the simplest mixture, two binodals can exist: a nematic-isotropic (N-I) binodal and an isotropic-isotropic (I-I) binodal. As this two curves depend on different physical parameters, they relative location can vary from one mixture to another; the I-I binodal might be placed at high temperatures and can be observed in equilibrium (Fig. 1b), or it can be buried bellow the N-I binodal at lower temperatures, so that the equilibrium phase diagram only shows I-N coexistence (Fig 1a). In this case, the I-I coexistence represents a metastable state.

Spinodal lines represent the limit of stability of homogeneous phases with respect to molecular fluctuations. The spinodal condition is met when an infinitesimal variation of a given variable produces a decrease in the free energy (a strict discussion on the spinodal criterion can be found in refs. [12;49]). Several spinodals can be defined, corresponding to different instabilities. The I-I spinodal represents the stability of a homogeneous isotropic phase respect to phase separation, so inside this region an isotropic phase is unstable with respect to concentration fluctuations. The I-I spinodal line is given by the condition $\partial^2 f / \partial \phi^2 = 0$. The nematic spinodal is the limit of stability of a homogeneous nematic phase, with respect to fluctuations in order or composition (This spinodal is often considered to extend up to its intersection with the NIT line, but actually it extend to higher temperatures. For the correct calculation of this line see refs. [12;49]). It is given by the conditions $\partial^2 f / \partial \phi^2 \, \partial^2 f / \partial S^2 - \partial^2 f / \partial S \partial \phi = 0$. Finally, another spinodal can be defined, the isotropic-to-nematic spinodal which is the limit of stability of a homogeneous isotropic phase with respect to fluctuations of the order parameter and it is defined by the condition $\partial^2 f / \partial S^2 = 0$, evaluated with $S = 0$. In previous works, usually the NIT line was considered to be stability limit of an isotropic phase respect to order fluctuations, but between the isotropic-to-nematic spinodal and the NIT line, $\partial^2 f / \partial S^2 > 0$ and $S = 0$ represents a local minimum of the free energy, so the isotropic phase is metastable (this is very well known for pure LCs but it was usually ignored for mixtures). These spinodals are schematically shown in Figs. 1a and 1b by dotted lines.

Figure 1 shows schematics of phase diagrams of mixtures containing a nematic LC and a non-nematogenic species including binodals and spinodals. I and N indicate isotropic and nematic phases. Figure 1(a) correspond to highly a miscible mixture, with buried (metastable) I-I phase coexistence, and Figure 1(b) to the case of low misciblitily with demixing in the isotropic phase. Characteristic positions corresponding to different quenches leading to different spinodal processes are indicated as a, b and c and discussed below in section 3.b.

## 2.1. Polymer – liquid crystal mixtures.

Phase diagrams of different types of PDLCs have been studied by several groups [10-22]. Low-molecular weight LC mixed with thermoplastic (i.e. not crosslinked) polymers, are described by Maier-Saupe / Flory Huggins theories, which gives the type of phase diagram shown in fig. 1. The dimensionless free energy density is given by [11-15;19;20]:

$$\frac{v_{ref} f}{RT} = \frac{\phi}{v_{LC}} \ln(\phi) + \frac{1-\phi}{v_P} \ln(1-\phi) + \chi\phi(1-\phi) + \frac{1}{2}\frac{v}{r_{LC}}\phi^2 S^2 - \frac{\phi}{r_{LC}} \ln(Z) \qquad (2)$$

where $v_{ref}$ is an arbitrary reference volume, $v_{LC}$ and $v_P$ are the molar volumes of LC and polymer, (non-dimensionalized with respect to $v_{ref}$), $\chi$ is the Flory interaction parameter, $v = 4.54\, T_{NI}/T$ is the Maier-Saupe quadrupolar parameter and $Z = \int_0^1 \exp\left[\frac{1}{2}\Gamma\phi S(x^2-1)\right]dx$ is the partition function. The first three terms correspond to the isotropic contribution ($f_{iso}$) and the last two terms are the nematic free energy ($f_n$). The main variables controlling the phase diagram are the interaction parameter and the relative volumes of the species. For highly compatible or low molecular weight species, the mixture is miscible and a phase diagram like Fig 1a is observed, as the molecular weight or the chemical incompatibility increase, a phase diagram like Fig 1b is observed. In addition, as the molecular weight of the LC increases, $T_{NI}$ increases. For large enough molecular weight, excluded volume effects becomes predominant, in these conditions a "chimney" can appear in the phase diagram, as observed in fig 2 [17;49]. This means that, for a concentration of LC higher that certain value, the system is in a nematic phase at every temperature. This can be represented by adding an excluded volume term to the quadrupolar interaction, $\nu = a_n + b_n/T$, where $a_n$ and $b_n$ are constants arising from excluded-volume and energetic interactions respectively. In Onsager's theory, $a_n = 5/4 \, L/D$, where L and D are the length and diameter of the LC molecule [10;17].

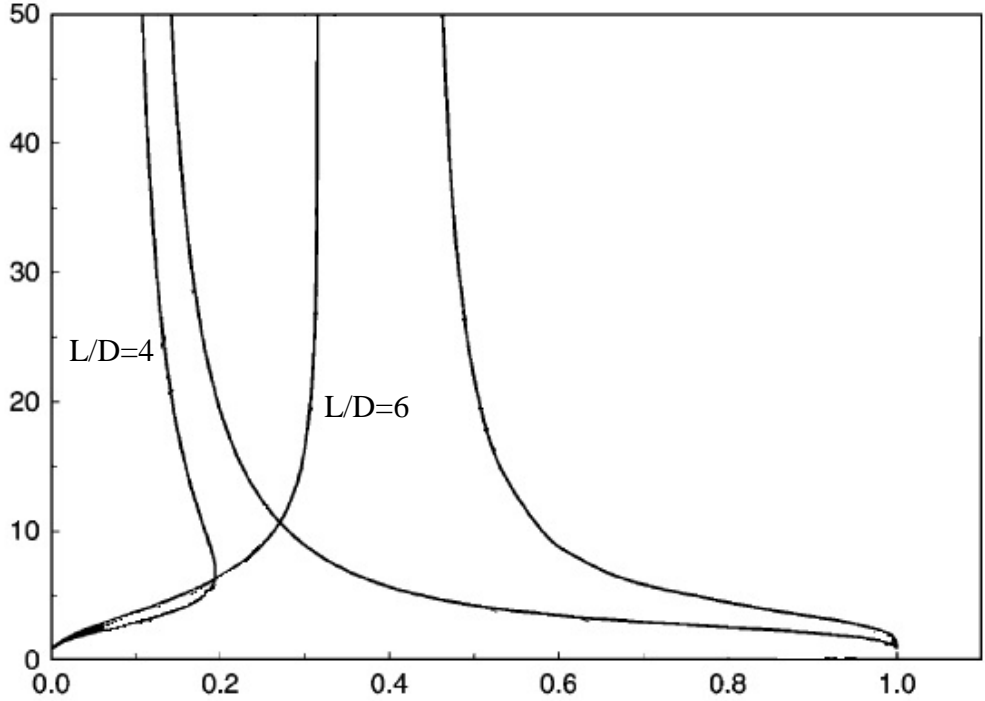

**Figure 2.** Phase diagrams of mixtures of a flexible polymer and a "long" liquid crystal. The dimensionless temperature is defined as $\tau=1/\chi$. The phase diagram is calculated with $v_P = 2$, $b_n=2.5$, and $v_{LC}=L/D= 4$ and 6 as shown in the figure. Adapted from ref [17], with permission of Elsevier.

In the case of a cross-linked polymer, the second term in eq 1, which represents the translational entropy of polymer chains, is absent, and in addition the elastic energy of the polymer network has to be taken added to the free energy. Several expressions have been used for the elastic energy in the literature and discussing them is beyond the scope of this review, a simple and classical expression based in Flory-Rehner theory of rubber elasticity is [11;13;14;50]:

$$\frac{v_{ref} f_{el}}{RT} = n_e \frac{3}{2} A (1-\phi_0)^{2/3} \left[ (1-\phi)^{2/3} -1 \right] + n_e B \ln(1-\phi) \qquad (3)$$

where $n_e$ is the number of elastically active chains, 1-$\phi_0$ is the polymer volume fraction when the network is formed (it can be different from 1-$\phi$ if the network is produced in-situ during a polymerization process, for example), and A and B are functions of the concentration and the functionality of the cross-links. The simplest biphasic equilibrium in this case consist in a swollen gel and pure LC (in principle, equillibrium between two gels with different degree of swelling or nematic gels can also exist). In this case, there is always a "chimney" in the phase diagram, as the gel cannot be infinitely swollen at any temperature. Benmouna *et. al.* analyzed [11;13;14] the phase diagrams predicted by different models for the elastic free energy. A representative phase diagram for this type of system is shown in fig. 3.

Das and Rey analyzed the computational aspects of phase diagram calculation [17], specifically the numerical accuracy of different ways of calculating the nematic partition function term in Maier-Saupe's energy. They analyzed three different approaches: A Landau-de Genes (LdG) fourth-order polynomial expression, based in a Taylor expansion of the integral, and two numerical integration schemes: Gaussian quadrature and Simpson`s rule. A LdG expression is attractive from a computational point of view, as it is simple to implement and requires less calculations than other techniques. But the Taylor expansion was also shown to be the less accurate approach. They concluded that a fourth degree Taylor expansion was very inaccurate, and Gaussian integrations must be performed with about 30 points to give a result comparable to the more accurate Simpson's rule. Later, Soule and Rey [51] proposed another strategy to obtain a highly accurate LdG expression: not a Taylor expression, but a least-square fit of the free energy with a polynomial of the desired degree. It was shown that a very accurate reproduction of the free energy and the phase diagrams can be obtained by using polynomials of degree four or five in a relatively wide range of temperature.

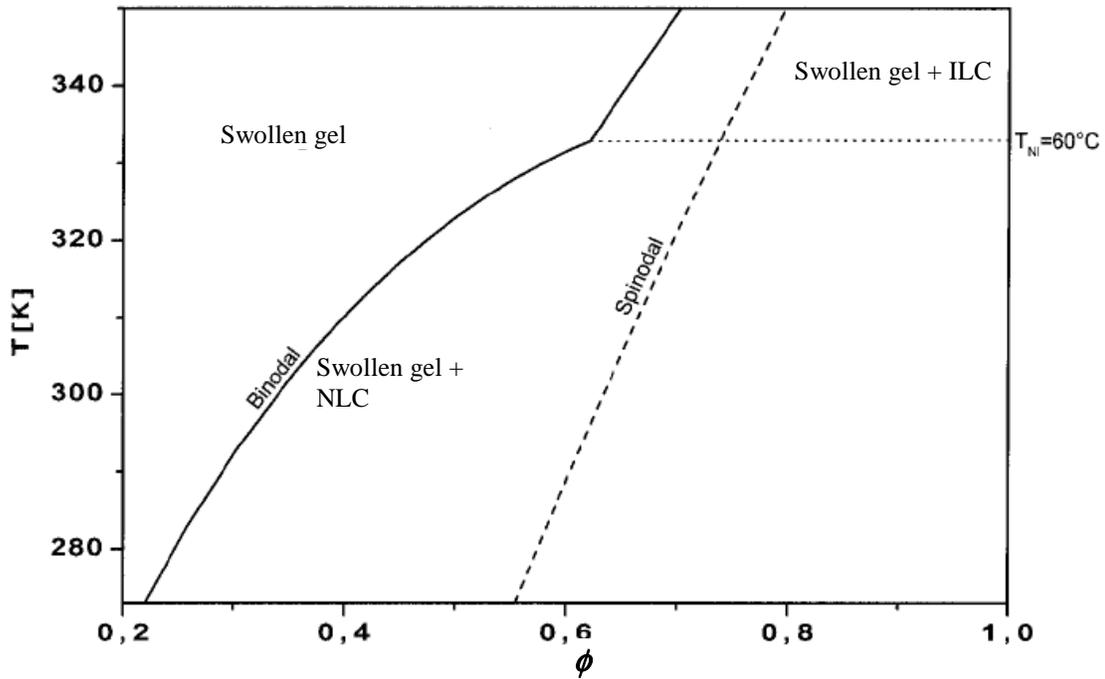

**Figure 3.** Phase diagram of a mixture of nematic liquid and a cross-linked polymer. NLC and ILC denote the pure LC in nematic or isotropic state, swollen gel is the polymer network swollen by the isotropic LC. The horizontal dotted line is the N-I transition of the pure LC phase. Adapted from ref. [13], with permission of John Wiley and sons.

**2.2. Liquid Cristal – Liquid Crystal mixtures**

In a mixture between two LCs, there are two isotropic-nematic transitions, each one corresponding to each of the liquid crystals [14;52-55]. The resulting nematic phase in the mixture of components "1" and "2" can be such that $S_1>S_2$ (phase $N_1$), $S_2>S_1$ (phase $N_2$) or $S_1=S_2$. This is controlled by the interactions between the two components and $T_{NI}$ asymmetry [54;55]. For ideal mixtures, the NIT temperature of the mixture is a linear interpolation of $T_{NI,1}$ and $T_{NI,2}$ while for strong deviations of ideality, azeotropic/eutectic behavior can be observed [52;54;55]. In general, the phase $N_1$ exist for compositions between the azeotrope and pure "1", $N_2$ between the azeotrope and pure "2", and for the azeotropic composition, $S_1=S_2$. For ideal mixtures, the component with higher $T_{NI}$ will have a higher $S$.

Maier Saupe theory can be applied to describe a nematic mixture, the nematic contribution to the free energy is [14;52-55]:

$$\frac{v_{ref} f_n}{RT} = \frac{1}{2}\frac{v_1}{r_1}\phi_1^2 S_1^2 + \frac{1}{2}\frac{v_2}{r_2}\phi_2^2 S_2^2 + \frac{1}{2}\frac{v_{12}}{\sqrt{r_1 r_2}}\phi_1\phi_2 S_1 S_2 - \frac{\phi_1}{r_1}\ln(Z_1) - \frac{\phi_2}{r_2}\ln(Z_2) \quad (4)$$

where $v_1=b_{n1}/T$, $v_2=b_{n2}/T$ and $v_{12}=b_{n12}/T$. This scalar formulation was first introduced by Brochard *et. al.* [52], recently Golmohamadi and Rey [54;55] derived the tensorial equations, which reduce to the scalar equation for a uniaxial phase in equilibrium.

Brochard *et. al.* first [52], and later others [14;53-55], described phase diagrams with complex shapes, including not only I-N and I-I phase coexistence but also N-N coexistence, both with maximum and minimum temperature azeotropes for generic mixtures. Some of them are shown in figure 4.

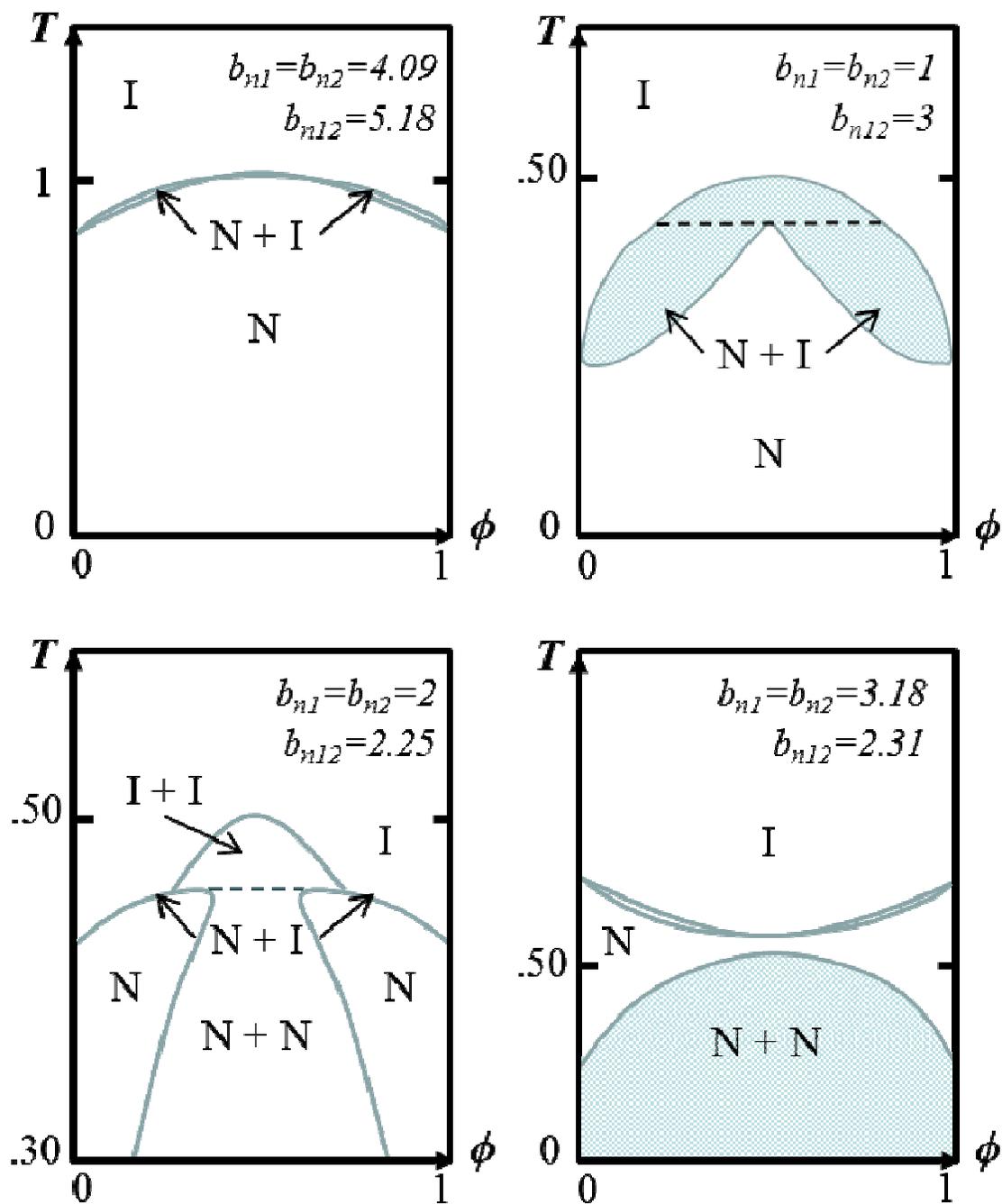

**Figure 4.** Some phase diagrams of LC – LC mixtures, reproduced from ref. [52] with permission of EDP sciences. All the phase diagrams were constructed with χ=1/T, and the nematic quadrupolar parameters indicated in each figure

In a recent work, Golmohamadi and Rey [54;55] focused on modeling the structure and properties of the nematic phase, applying the model to carbonaceous nematic mesophases, which consist in mixtures of discotic molecules with similar chemistry and different molecular weight. They combined the theoretical model with experimental information for $T_{NI}$ as a function of molecular weight (recall that higher molecular weight is associated with a higher $T_{NI}$). An important result was the quantitative classification of the behavior of the mixture in terms of the value of the interaction parameter (defined as $\beta=b_{n12}/b_{n2}$) and the molecular weight difference $\Delta M$, which is shown in Fig. 5.

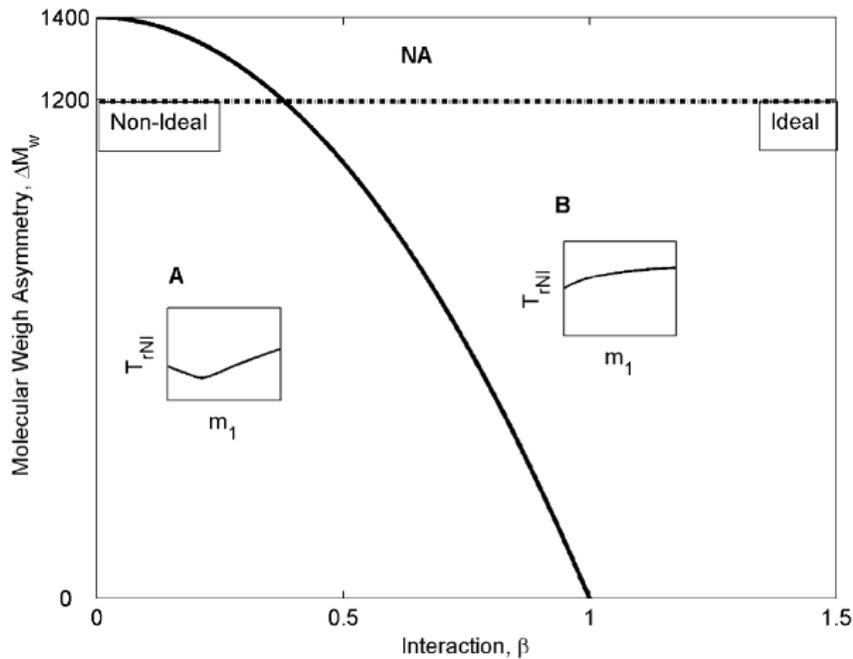

**Figure 5.** classification of an LC-LC mixture in terms of ideality, as a function of the interaction parameter $\beta$ and the molecular weight asymmetry. Adapted from ref. [55].

A second important result was the derivation of an analytical expression for the azeotropic composition:

$$\beta = \frac{(1-m_{1,c})\left(1-\frac{\Delta M}{M_1}\right)\frac{T_{NI,2}}{T_{NI,1}} - m_{1,c}}{(1-2m_{2,c})\sqrt{1-\frac{\Delta M}{M_1}\frac{T_{NI,2}}{T_{NI,1}}}} \qquad (5)$$

where $M_1$ is the molecular weight of "1", $m_{1,c}$ and $m_{2,c}$ are the mass fraction of "1" or "2" at the azeotrope. With this equation the interaction parameter can be determined by measuring the azeotropic composition.

**2.3. Dispersions of Nanoparticles in Liquid crystals**

Phase diagrams for dispersions of micron-colloidal particles and nematic liquid crystals have been studied in the past by several authors [26;28;30;32;56-58]. A colloidal particle embedded in a nematic phase induce elastic distortions in the director field and the formation of topological defects [23;24;59-62], which increases the total free energy producing a decrease in the NIT temperature. In a continuous solution-thermodynamic formalism [28;56;58], this effect is introduced as an interaction term, proportional to $\phi(1-\phi)S^2$. In addition, the formation of ordered arrays of particles is dictated also by elastic effects. In this case of micron-sized particles (much larger than single molecules), elastic effects dominate the free energy. For nano-sized particles, the situation is different in that the size of the particles is comparable to the size of a molecule, so mixing entropy and entropically-driven hard-sphere crystallization become relevant. Theoretical studies of phase diagrams for NPLC have been recently presented [26;28;58]. Simple models consider two first-order transitions: nematic ordering of the LC and colloidal crystallization of the particles. Such a model was first presented by Matsuyama and Hirashima [28], and later modified by Soule *et. al.* [58], who considered the following modifications. First, as the particles are not flexible chains, Flory-Huggins theory has to be modified. This is done by introducing an excluded-volume term, obtained from Carnahan-Starling equation of state. This idea was first proposed by Ginzburg for polymer nanocomposites [63], and has been widely used since then [26;64;65]. A second modification is considering that the interactions are proportional to the contact areas, and not to the volumes of the components [58;65]. By considering that the nematic quadrupolar interaction is also proportional to the contact area, a consistent dependence of the isotropic-nematic transition with particle radius is reproduced (i.e. as the particle radius increase, $T_{NI}$ of the mixture increase, approaching the value of the pure liquid crystal in the limit of infinitely large particles). The free energy for this system is written as the summation of four contributions, isotropic ($f_{iso}$), nematic ordering ($f_n$), crystalline ordering ($f_{crys}$) and specific interactions ($f_{int}$) [26;28;58].

$$\frac{v_{ref} f}{RT} = f_{iso} + f_n + f_c + f_{int} \qquad (6)$$

where:

$$f_{iso} = \frac{\phi_{LC}}{v_{LC}} \ln(\phi_{LC}) + \frac{\phi_{NP}}{v_{NP}} \ln(\phi_{NP}) + \frac{\phi_{NP}}{v_{NP}} \frac{(4\phi_{NP} - 3\phi_{NP}^2)}{(1-\phi_{NP})^2} + \chi a_{NP} \phi_{NP} \varphi_{LC} \qquad (7)$$

$$f_{nem} = \frac{\phi_{LC}}{v_{LC}} \left[ \frac{1}{2} v \varphi_{LC} S - \ln(Z) \right] \qquad (8)$$

$$f_{cris} = \frac{\phi_{NP}}{v_{NP}} \left[ \frac{1}{2} g \phi_P \sigma^2 - \ln Z_P \right] \qquad (9)$$

$$f_{int} = w S^2 a_P \phi_P \varphi_{LC} + c S \sigma a_P \phi_P \varphi_{LC} \qquad (10)$$

The third term in eq 7 is the Carnahan Starling contribution, $\varphi_{LC} = \phi_{LC}a_{LC}/(\phi_{LC}a_{LC} + \phi_{NP}a_{NP})$ is the area fraction of liquid crystal, $a_{NP}$ and $a_{LC}$ are the area per unit volume of the particle and liquid crystal, $\sigma$ is the crystal order parameter, $Z_P$ is the positional partition function of the particles, $g$ is an excluded-volume parameter (=14.95 for hard spheres), $w$ is a binary nematic interaction parameter that account for anchoring at NP surface and nano-scale disruption of order, and $c$ is a crystal-nematic coupling parameter, which phenomenologically account for the fact that an ordered array of particles can be favored by elastic forces in a nematic matrix [28].

Figure 6 shows some representative phase diagrams calculated by Soule *et. al.*, for different NP radius. Miscibility is a balance between the entropic and enthalpic terms, which depend on the particle radius in different way. For small particles, increasing the size decreases miscibility(entropic effect prevail), while for large particles the opposite happens (enthalpic effect) [64;65]

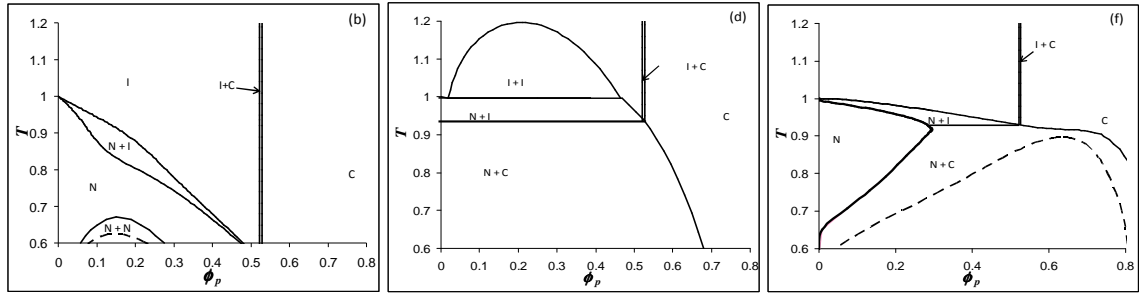

**Figure 6.** Calculated phase diagrams for NP-LC mixtures, as given by eqs , with the following parameters: $v_{LC} = 3$, $a_{LC} = 4.66$, $\chi = 2.5/T$, $w = c = 0$, and different values of NP radius: (a) $R_{NP} = 0.9$, (b) $R_{NP} = 2$, (c) $R_{NP} = 6$ (here $v_{NP} = 4/3\pi R_{NP}^3$ and $a_{NP} = 4\pi R_{NP}^2$). Adapted from ref. [58].

This model was later extended to system consisting in functionalized NPs [32], where the metallic core of the particle is coated by a mixture of two different ligands: a long ligand with a messogenic group, and a short alkylic ligand. A phenomenological expression for the interaction parameter $\chi$ was introduced for the entropic and enthalpic effects produced by the partial penetration of LC molecules in the corona of ligands, taking into account that the two ligands have a different size. This situation is schematically shown in figure 7.

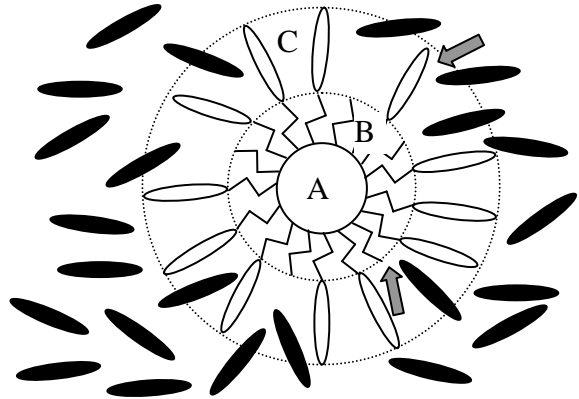

**Figure 7.** Schematic of a particle coated with a mixed ligand layer, in presence of LC solvent (black ellipsoids). Region A is the gold core, Region B is the inner layer composed of alkyl chains (zigzag lines), and Region C is the outer layer composed by the mesogenic group of the ligands (white ellipsoids). The partial penetration of the solvent molecules into the outer layer of the corona is shown. Examples of LC solvent molecule in direct contact with an alkanethiol chain and with a liquid-crystalline ligand are indicated by the grey arrows.

The proposed expression for the interaction parameter as a function of the fraction of long ligand was:

$$\chi = A_0\left[(\theta-0.5)^2 - 0.25\right] + \frac{B_1\theta + B_2(1-\theta)}{T} \tag{11}$$

The first term is the entropic contribution and it a symmetric parabola with a minimum at θ=0.5. This terms takes into account that the penetration of LC molecules in the corona is maximum (so the entropy is minimum) for the case of mixed ligands, and it is negligible (the entropy was taken as 0) for coronas of pure ligands where there is no free volume. The second term is the enthalpic contribution and it is a linear function of the number of contacts between solvent-alkanethiol ligand and solvent-liquid crystalline ligands (the number of contacts with each ligand was considered to be proportional to θ and 1-θ). This model was able to reproduce a non-trivial miscibility trend for a series of nanoparticles, showing maximum solubility for θ close to 0.5, partial solubility for θ=1 and negligible solubitily for θ=0.

## 3 – Dyamics of phase transitions

There are many approaches to modeling the dynamics of phase transitions. In the continuum formulation, based in macroscopic thermodynamic variables, an equation of change is formulated for the order parameters and the concentration. These equations can be complemented with some equation for the variation of pressure (in compressible systems), temperature and velocity.

Hohenbergh and Halpering [66] classified the dynamic models in terms of the relevant variables describing the process. A system with only non-conserved order parameters is known as model A and it is described by the following dynamic law:

$$\frac{\partial \psi}{\partial t} = -M_\psi \frac{\delta f}{\delta \psi} = -M_\psi \left( \frac{\partial f}{\partial \psi} - \nabla \frac{\partial f}{\partial \nabla \psi} \right) \qquad (12)$$

where Ψ is the order parameter and δ represents the functional derivative. Here and bellow, $M_i$ represents a mobility corresponding to variable "i". In order to describe nematic order, the tensorial formulation is used and the model A equation becomes:

$$\frac{\partial \mathbf{Q}}{\partial t} = -M_Q \left( \frac{\partial f}{\partial \mathbf{Q}} - \nabla \cdot \frac{\partial f}{\partial \nabla \mathbf{Q}} \right) \qquad (13)$$

This equation is phenomenological in principle; is the simplest equation that guarantees that the entropy of the system decreases monotonically with time, although there have been several works relating this equation with more fundamental laws from non-equilibrium thermodynamics or molecular theories [67-69].

A model described by a conserved variable is known as model B and it is described by Cahn-Hilliard equation:

$$\frac{\partial \phi}{\partial t} = \nabla \left[ M_\phi \nabla \left( \frac{\partial f}{\partial \phi} - \nabla \frac{\partial f}{\partial \nabla \phi} \right) \right] \qquad (14)$$

In this case, as there is an extra restriction (conservation law), the minimum equation is different. Cahn-Hilliard equation can be though as a generalization of Fick's law and can be derived from Onsager's irreversible thermodynamics or other formalisms [67-69].

In the case of a mixture undergoing an order-disorder transition, the minimum model has to account for ordering and concentration and thus requires one non-conserved and one conserved order parameters, this is known as model C. More complex formulations can include an equation for energy and momentum transfer.

### 3.1. Interfacial kinetics - Metastable fronts

The growth of a nematic spherulite can be represented by the interface normal velocity $v_I$, which it is experimentally found to follows a power law with time, $v_I = dR/dt = a.t^n$, where $a$ is a constant (related to the driving force), and the exponent $n$ depends of the type of phase transition process ($n=0.5$ for diffusional and $n=1$ for non-difussional transformations). For the case of a pure liquid crystal, it is observed that $n=1$ for large undercooling, while $n$ approaches 0.5 when the temperature is close to the bulk equilibrium transition conditions [70-72]. Traditionally this was ascribed to the effects of interfacial energy (the argument was that, as the transition temperature is approached, interfacial energy dominates over bulk energy and the process is driven by interfacial dynamics), but recently it was shown that it is not the interface, but the latent heat released by the transition which produces this $n < 1$ [73;74]. The coupling between a non-diffusional process (phase ordering) and a diffusional one (heat transfer), lead to a complex dynamic behaviour.

As mentioned before a first order transition in a mixture is described by model C, which couples a difussional and a non-diffusional dynamic equation corresponding to the non-conserved and the conserved variables respectively. If one variable is much faster than the other, then the dynamics will be controlled by the slower variable and will show $n=0.5$ or $n=1$ depending on the case. When the two variables have a comparable dynamics, then the value of $n$ is intermediate. A characteristic feature of this mixed process is that, as the diffusional velocity decreases with time while the ordering velocity remains constant and the system is controlled by the slower variable, at long enough times the system will be always controlled by diffusion [75;76]. So, even when ordering is slower at first and the initial growth exponent approaches 1, it will decrease with time and approach the diffusive value of 0.5.

Another interesting feature of phase transformations is the possibility of formation of metastable states. For example, as discussed in the previous section, I-I coexistence can be buried bellow I-N equilibrium (fig.1a), and thus be metastable. If an isotropic phase is quenched to a large undercooling, the presence of this metastable equilibrium can affect the dynamics of the system and the metastable phase can be formed through different mechanisms. Bechoeffer *et. al.* [77;78] first found, for a model of non-conserved order parameters, that an interface separating the stable phases can spontaneously split in two, so a third phase (the metastable one) is formed. Later, Evans *et. al*. [79-81] analyzed the case of a conserved order parameter (COP). More recently, Soule and Rey [75;76;82] analyzed the case of mixed order parameters, and found a complex behavior, arising from the more complex dynamics and phase behavior.

There are two main factors controlling the structure of the interfaces appearing during the dynamic process: 1 – the relative location of the initial condition with respect to the metastable I-I coexistence curve in the phase diagram, 2 – the relative mobility. The position in the phase diagram determines the possible mechanisms available for the appearance of a metastable phase [82]. This is shown in figure 8 and will be discussed in the following paragraphs.

In the phase diagram indicated in figure 8, four characteristic phases, corresponding to the stable and metastable equillibria, can be found. These are indicated in figure 8 with greek letters as follows: $\alpha$ and $\beta$ correspond to the stable-equillibrium nematic and isotropic phases respectively, $\delta$ and $\gamma$ correspond to the metastable I-I equilibrium, being δ the low-concentration and γ the high-concentration phases. If a nematic nucleus is formed in an isotropic media in $(T,\phi)$ conditions comprised inside the I-N binodal, the LC nematic phase will grow at the expense of the isotropic one. In addition, when the concentration of LC is lower than the I-I critical concentration, and the isotropic phase is inside the I-I binodal, the concentrated isotropic phase γ, can grow at the expense of the bulk isotropic phase. In this case, the interface spontaneously splits in two: one is a α–γ interface and the second one is a γ–δ interface, [75;76]. This splitting is a kinetic mechanism that takes place when the γ–δ interface is initially faster than the α–γ interface. As the γ–δ interface separates two I phases, it follows a diffusional kinetics ($v_I = a.t^{-1/2}$), while the α–γ interface separates I and N phases and follows a NCOP kinetics ($v_I = const$), after some time the interfaces merge again and the split state has a finite lifetime.

When the concentration of LC is higher than the I-I critical concentration, the situation is different. When ordering is faster that diffusion, the kinetic front splitting is not observed (as in this case, γ cannot grow at the expense of the bulk). Nevertheless, if the interface is at equilibrium, (diffusional transformation) the concentration of the isotropic phase at the interface is the equilibrium one (β), a depletion layer is formed between the interface and the bulk, and the concentration of this depletion layer goes through the I-I binodal. Under this condition, a phase separation takes place within the depletion layer [82]. Unlike the previous case, this is a thermodynamic and not a kinetic effect. As both interfaces follow a difussional kinetics, the separation between interfaces grows as $v_I = a.t^{-1/2}$ and the split state has an infinite lifetime. The double front in this case arises even when the initial condition is outside the I-I binodal, the condition for it to happen is that the temperature is lower than the I-I critical temperature. In addition, even above the critical temperature, diffusion shows a strong non-classical behaviour: an inflection point is produced in the concentration profile in the depletion layer when the temperature is close to the I-I critical temperature. This is ascribed to the fact that the second derivative of the free energy with respect to concentration goes close to 0 in the vicinity of the critical point [82].

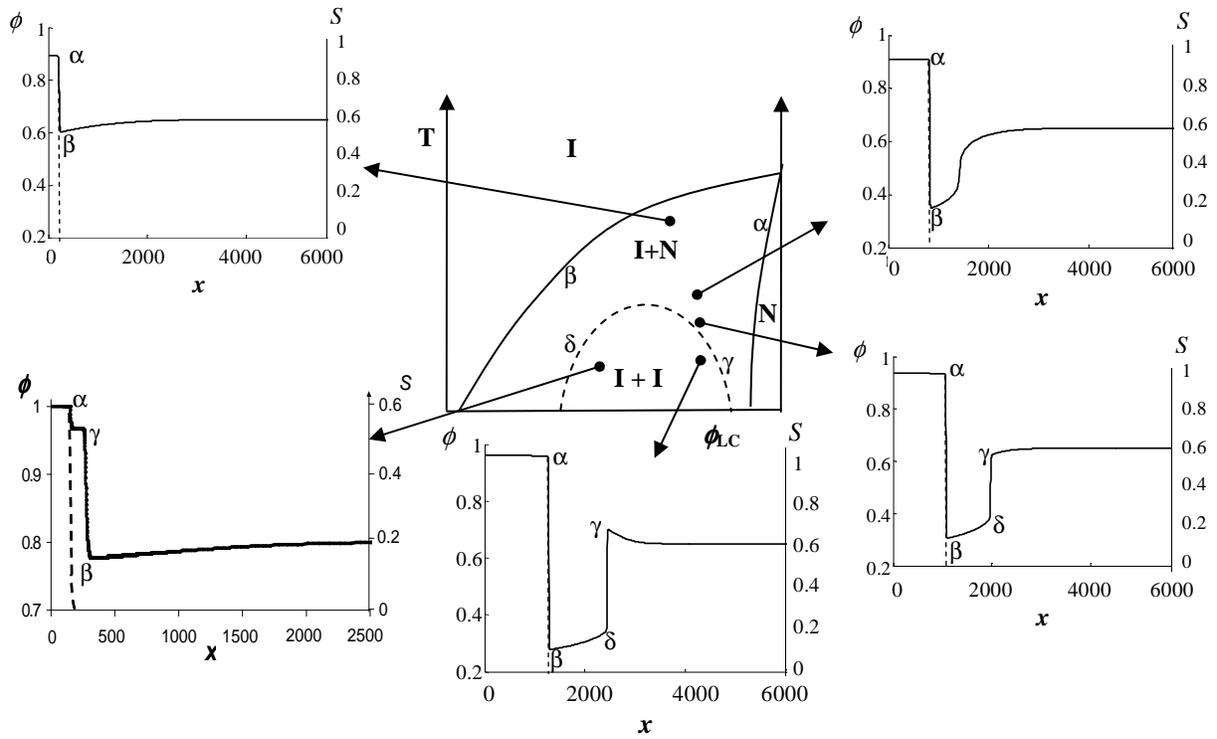

**Figure 8.** Interface splitting in different locations of the phase diagrams. The central plot is a schematic phase diagrams, where the solid lines represent the stable I-N binodal and the dashed line is the metaestable I-I binodal. The satellite plots are the profiles of concentration (solid lines) and order parameter (dashed lines), obtained from simulations in the region of the phase diagram indicated by the corresponding arrows. Greek letters in the phase diagrams and in the profiles indicate the different phases ($\alpha$: stable nematic, $\beta$:

stable isotropic, δ and γ: metastable isotropic). The concentration of the bulk isotropic phase is represented by black dots. Adapted from refs. [75;76;82], with permission of EDP sciences.

This situation is graphically shown in figure 8, where the different composition and order parameter profiles (taken from simulations) are shown for different locations in the phase diagram (shown schematically).

### 3.2. Spinodal decomposition

Different spinodal lines can be defined in mixtures involving LC, as discussed previously. Each spinodal line is associated with a specific phase transition process. Fiescher and Dietrich [83] first studied spinodal decomposition for a mixed order parameter case where only phase separation between ordered and disordered phases was possible. Soule and Rey [76] later extended it to the case where a metastable I-I phase separation is also possible. Das and Rey [18;21;22] also performed 2D simulations of spinodal decomposition, analysing the phase transition dynamics following quenches to different locations in the phase diagram. Different dynamic regimes are observed depending on the location in the phase diagram and the relative mobilities. When the system is quenched to point b in figure 1b, (where it is stable with respect to composition fluctuations but unstable respect to order fluctuations), the system first becomes ordered, and then it phase-separates to the equilibrium state (due to order and composition couplings). If it is quenched to point a (unstable respect to concentration, stable respect to order), it first phase separates into two isotropic phases, and then the concentrated phase phase separates again into a ordered and a disordered phase. The time interval between the two steps will depend on the relative mobility. If it quenched to point c (unstable respect to order and concentrations), different regimes are possible depending on the relative mobilities: the system can become ordered at homogeneous concentration and then phase separate, it can phase separate (to the metastable I-I phases), and then undergo a secondary I-N phase separation, or phase separation and ordering can evolve simultaneously and fully coupled [76].

**4 – Structure: morphologies, textures and defects.**

As discussed before, different complex dynamic processes are available, depending on the location in the phase diagram and the relative mobilities. Not only the kinetics of these processes is complex, but also a rich variety of morphologies and structures can be formed. In addition, the final structure in the ordered phase (director configuration, defects, etc), depends not only on the thermodynamics but also on the boundary conditions (anchoring).

**4.1. Domain Morphologies in phase separation**

Several different morphologies can be produced by a phase transition, depending on the $(T,\phi)$ initial conditions and the kinetic parameters. It has been shown in the previous section how double-fronts (core-shell structures) can be formed in a nucleation and growth process for certain conditions, and that in spinodal decomposition the process can be step-wise, which affects not only the kinetics but also the domain morphology. For example, 1D simulations show that if I-I phase separation precedes ordering, a salami-type structure can be formed, where big domains of high concentration formed by smaller subdomains of ordered phase are formed [76].

Das and Rey preformed 2D simulations for a PDLC in different locations in the phase diagrams[18;21;22]. The main results are shown in figure 9. Fig 8a shows the case where the system is quenched to a region of the phase diagram where it is initially unstable respect to phase separation and metastable respect to ordering (point a in figure 1). As the concentration of LC is higher than the I-I critical concentration, LC rich domains are formed as a dispersed phase, although partially interconnected. As the concentration in the LC-rich domains increase, the domains become ordered forming the nematic phase. They did not observe a clear secondary phase separation leading to salami structures for the kinetic parameters used, but they did observe a breaking down of interconnected domains into single droplets when the domains became nematic. Fig 8b correspond to the case where the system is initially unstable respect to ordering and metastable respect to phase separation (point b in figure 1). In this case, the system first orders and then phase separates, so disperse isotropic domains are expelled from the continuous nematic matrix. Fig 8c correspond to the case where the system is unstable respect to both ordering and phase separation (point c in figure 1), as shown before the dynamics and the structure in these conditions will strongly depend on the relative mobilities, in the case analyzed by Das and Rey (where ordering was relatively fast), the structure is similar to case b, except that the isotropic domains are highly interconnected. Under certain conditions, the dispersed isotropic droplets in a nematic matrix can form ordered arrangements (see next section).

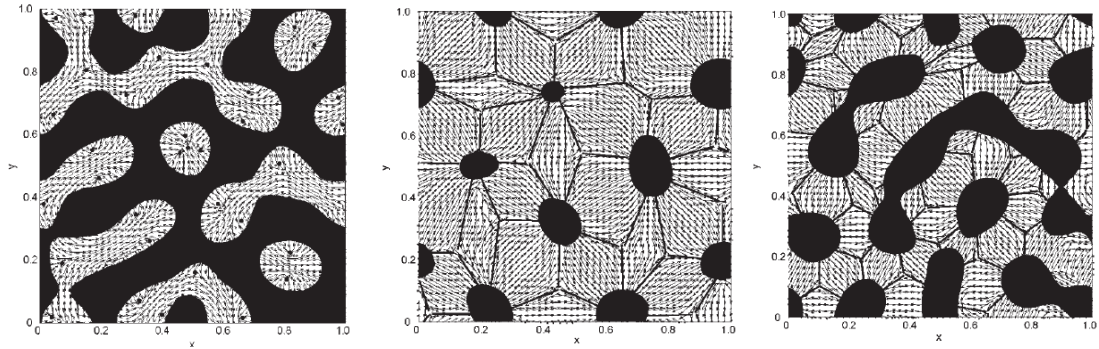

**Figure 9**. Snapshot of the local composition of the system at a late time step following a quench to; (a) point a, (b) point b, (c) point c, for a polymer-liquid crystal mixture. Black corresponds to isotropic polymer-rich and white corresponds to LC-rich phases. The arrows represent the local nematic director, and defects are marked with small solid circles. Reproduced with permission from ref. [18], copyright (2004) American Institute of Physics.

**4.2. Textures and defects in mixtures**

The study of defect structure and textures in a mixture is more complex than for a pure LC as the spatial variations of order and orientation are coupled with gradients in concentration, such that there can be preferential segregation of the species (for example, the non-liquid crystalline component tends to segregate preferentially to regions of lower order, like defect cores or boundaries between nematic domains), and this can not only modify the characteristics of some specific structure or configuration (e.g., modify the size of a defect), it can also modify the relative stability of different structures. This effect is important, for example, in the case of blue phases (which consist in a network of disclinations): the temperature range where this phase is stable can be increased from several degrees to a few dozens of degrees by adding a guest component [84-87].

Recently, Soule and Rey [88] analyzed how a hedgehog defect in a mixture of an LC and a isotropic guest component is affected by temperature and composition, extending a previous study by Mottram and Sluckin [89] for a disclination in a pure LC. The analysis was made by considering two complementary approaches: the continuum Landau-deGennes simulations, and a sharp-interface solution thermodynamic model (where the defect core is considered as an isotropic phase of a given radius, in equilibrium with the nematic bulk). It was found that the isotropic component segregates preferentially to the defect core, and the radius of the defect increases abruptly as the temperature and concentration approach the binodal line, and that a small range of supersaturation or superheating (where the nematic phase with the defect is metastable) is possible. Some profiles of order parameter $S$ and concentration of guest component $\phi_I$ are shown in figure 10, and the dependence of the defect radius on the global concentration of guest component (expressed as deviation from saturation concentration) are shown in figure 11.

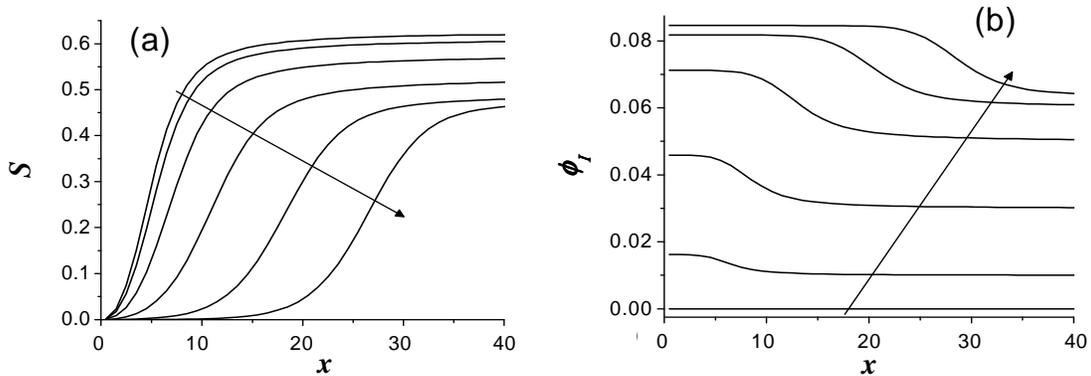

**Figure 10.** Profiles of order parameter (a) and concentration (b), for $T/T_{NI} = 0.925$ and the following values of $\phi_0$, increasing in the direction of the arrow: 0, 0.01, 0.03, 0.05,

0.06 and 0.063. (Note in b that for $\phi_0 = 0$, $\phi_1 = 0$). Adapted from ref. [88] with permission of The Royal Society of Chemistry.

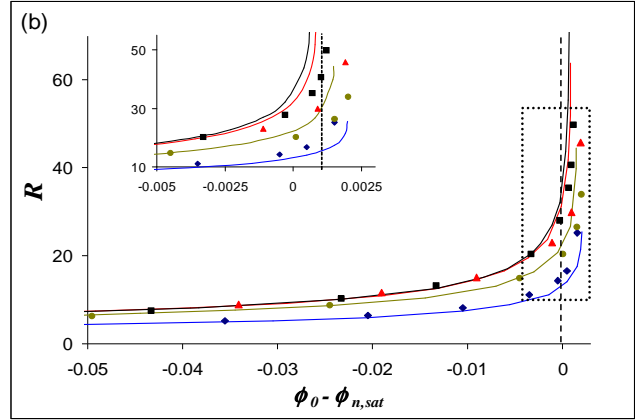

**Figure 11.** Defect core radius, as a function of the deviation of the global concentration from saturation conditions. The dashed line shows the saturation (binodal) composition. The inset corresponds to the area in the vicinity of saturation, as indicated by the dotted square. The symbols are the result from LdG simulations, the full lines are the results from the analytical theory. The different temperatures are $T = 0.925$ (squares), 0.85 (triangles), 0.775 (circles), and 0.7 (diamonds). Adapted from ref. [88].

When a higher dimensionality is considered in a phase separation process, different defect structures can arise due to the anchoring conditions imposed by the nematic-isotropic interfaces. 2D simulations by Das and Rey [18] showed that the, when the nematic phase is present as dispersed domains, several +/-1/2 defects were formed, favouring the formation of a bipolar structure in spherical droplets, as shown in fig 8a. When the nematic phase is continuous, a structure of polygonal domains of nematic phases with isotropic droplet is observed [17]. They found that modifying the interface thickness (which is directly related to interfacial tension) lead to different regimes with corresponding different morphologies. For large interfacial tension, the droplets are much larger than the characteristic length of texturing and an ordered array consisting in networks of isotropic droplets and defects is formed, as shown in figure 12. This type of textures are similar to those that can be observed in suspensions of colloidal particles in nematic LCs, which are discussed in the next section. As the interfacial tension decreases, the droplet size decreases and there is a transition from an ordered array to a random dispersion of droplets. For very small interface tension, the droplet size becomes comparable to the thickness of the interface and much smaller than the characteristic texturing length scale, so order becomes frustrated and a very low value of order parameter, with a random orientation field is produced.

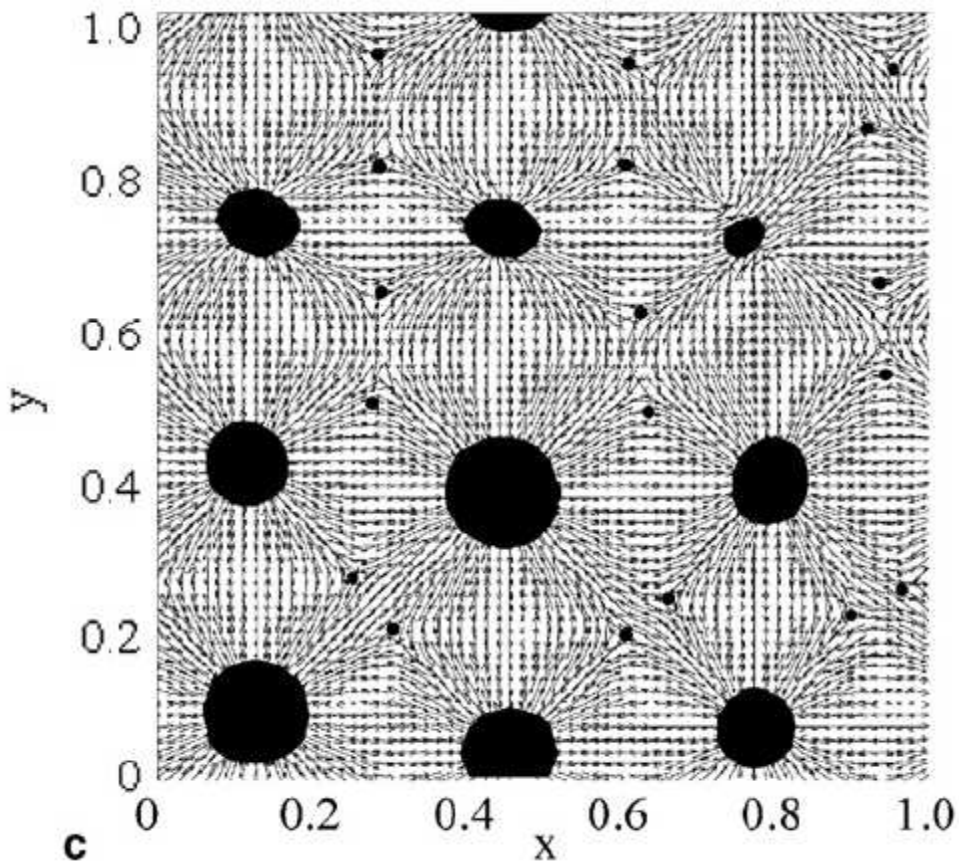

**Figure 12**. Ordered array of polymer droplets and defects in a nematic phase produced by phase separation, obtained from ref. [22], copyright 2006, with permission of Elsevier.

**4.3. Nano/micro - scale textures in filled nematics**

A filled nematic is similar to the structures shown in figure 12, where the "filling" are the droplets of the isotropic phase. In that case, the structure can be controlled by modifying the concentration of the mixture, the size and morphologies of the dispersed phase is self-selected. When solid colloidal particles are dispersed in the LC, the size, shape and concentration of the particles are imposed externally, in addition the anchoring at the surface of the particle can be tuned by surface treatments.

In principle, different approaches can be taken to describe filled nematics. A fully continuum, macroscopic formulation, would treat both the LC and the particles as continua, the mixture is described by a macroscopic concentration of particles, and the effect of particles on the nematic matrix are introduced in a mean-field approach as "interaction parameters". This approach is that described in section 2.3 for NP-LC mixtures. A second approach is a molecular formulation, where individual particles and individual LC molecules are considered, and described by molecular dynamics or Monte Carlo simulations. A third approach is an intermediate case, and consist in treating the LC as a continuum and the particles as individual entities (which act as boundary conditions for the LC). This is equivalent to molecular simulations with "implicit solvent", but in this case the solvent is structured and it is described by a complex dynamic equation (model A). This approach will be analysed in this section.

A complete description of this system requires a model for the evolution of the nematic matrix (model A), coupled with equations of movement for the particles (Brownian motion) [25; 29]. Nevertheless, a model with immobile particles can be useful to analyze the laws describing defect charges, defect configuration, texture transitions, interaction between the particles mediated by the nematic phase, etc. [24;59;60;90].

Gupta and Rey [61;62], and later Phillips and Rey [91;92], analyzed defect configuration for micron, sub-micron and nano sized spherical particles in polygonal arrangements. For the case of strong anchoring at the particle surface, the charge of the defects inside the polygon is given by Zimmer's rule: $C=-(N-2)/2$, where C is the total charge and N is the number of particles in the polygon. The defect structure and configuration was strongly dependent on the particle´s size. For micron particles, the defect structure depends on the number of particles forming the polygon; for an odd number, singular core defects of charge $-1/2$ are formed, such that the total number of defects satisfies $C=-(N-2)/2$, while for even number, a single escape-core defect was preferred [61;62]. For sub-micron particles and nanoparticles, only singular $-1/2$ defects were found [61;62;91]. In addition, for temperatures approaching $T_{NI}$, complex biaxial structures were observed, which were dependent on the boundary conditions.

When faceted NPs are dispersed in a nematic matrix, the geometric discontinuity propagates through the LC, and this can produce more complex structures than in the case of spherical inclusions [92-94]. As the topological charge of a surface defect $C_s$ on a faceted particle is just the ratio of the misorientation angle between two adjacent faces and $2\pi$, the defect an edge (3D) or a corner (2D) can absorb or emit is $\pm C_s$. For example for a square particle, where the relative angle between to adjacent phase is $\pi/2$, the surface defect charge associated with a corner is $C_s = \pm(\pi/2)/2\pi = \pm\pi/4$, where the sign depends on the director rotation when encircling the defect in a counter-clocwise direction.

It has been shown experimentally [95;96] that the interactions of nanoparticles in LC can be tailored by controlling the nanoparticle's shape, and different types of self-assembled structures can be obtained. Simulations for single particles [92;94;97] show that defects can be absorbed as surface defects in the corners of the particle, and pairs of defects can be linked through biaxial strings, as shown in figure 13. Neighbouring particles at small enough distance are linked through defect lines, while at large distances they behave as independent particles. Phillips *et. al.* [93] studied polygonal arrangements of faceted nanoparticles (in triangular, square, pentagonal and hexagonal geometry), and found that bulk defects are produced for small separations and low temperatures, and surface defects otherwise. An odd-even effect was found for this system too, where the insertion energy (excess free energy per particle) is higher for polygons of even number of sides [93].

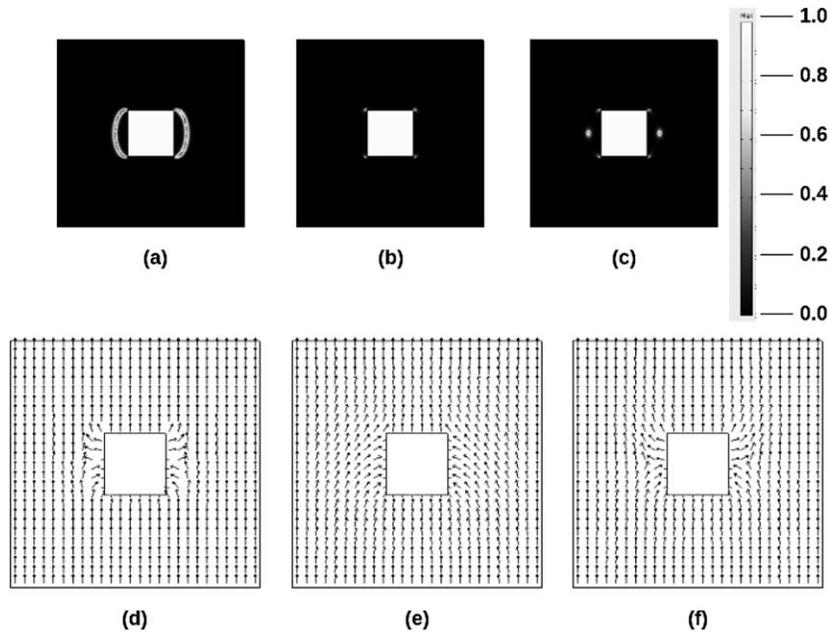

**Figure 13.** Computed gray scale visualization of the biaxiality parameter defined as $\beta= 1-6[(\mathbf{Q}.\mathbf{Q}):\mathbf{Q}]^2/(\mathbf{Q}:\mathbf{Q})^3$ showing the three defect modes : string mode (a), bulk and surface defect mode (b) and surface defect mode(c). The corresponding director fields associated with these modes are represented in (d), (e) and (f) respectively, for a temperature close to $T_{NI}$. Adapted from ref. [94] with permission of Cambridge University press..

When mobility of the particles is considered, self-assembly of the particles and macroscopic phase separation can be predicted, while the details of the textures at a nanoscopic level can be retained. Yamamoto *et. al.* [25] and Zhou *et. al.* [29] performed this type of simulations and observed the spontaneous formation of linear and bi-dimensional arrays of particles.

## 5. Conclusions

Liquid crystals are viscoelastic anisotropic soft matter materials, that combine the fluidity of liquids and the anisotropy of solids. They form the basis of many optical devices, sensor/actuators, drug delivery, structural fibers, and lubrication. Biological liquid crystals are found in membranes, DNA and protein solutions, and carbohydrates.

In many instances mxing between mesogens with non-mesogenic solvents and polymers, cross-linked macromolecules, colloidal and nanopartciles is used by man or Nature to improve performance, increase efficiency, facilitate processing, lower energy loads, and/or optimize material properties.

In other cases demixing through thermodynamic instabilities is used to create multi-phasis material architecture to achieve optical functionality (as in PDLCs) or mechanical strength (as in fiber re-inforced composites of LC polymers fibers embedded in a thermoplastic matrix).

In yet other cases, mesophase polydispersity is found naturally, as in carbonaceous mesophases from petroleum or coal pitches, resulting in precursors materials for high performance fibers consisting of molecules with significantly different molecular weight.

Hence accurate and reliable thermodynamic modelling continues to be at the forefront in developing new materials and devices as well as in providing a quantitative understanding of biological mesophase behaviour.

The present review provides a survey of the main thermodynamic theories, models, calculation methodologies for an important selection of mesogenic systems:

(i) monomeric mesogen and isotropic solvent

(ii) lyotropic liquid crystal polymer

(iii) monomeric mesogens and thermoplastic polymers

(iv) monomeric mesogen and cross-linkable monomers

(v) binary monomeric and mesogens

(vi) monomeric mesogens and colloidal and nanopartciles

Revealing specific features in the free energy of mixing that accounts for molecular details in the interaction parameters, and elastic and entropic contributions. A number of generic stability and metastability features in the various phase diagrams are highlighted to emphasize the novel aspects of phase transitions and phase separation in mesogen-containing mixtures.

Since the kinetics of transformations is a significant aspect of material fabrication, extensive discussions, analysis, and predictions on new growth laws are presented. The role of metastability, relative mobilities of order and diffusion, proximity of spinodal lines, shows that droplet growth may contain metastable coronas whose lifetime is affected by the above-mentioned effects. Recent integrated analysis and computations of metastable fronts in mesogenic mixtures extends previous work on mixed order parameter systems.

Leveraging thermodynamic instabilities and phase transition is a well establish path way to create morphologies and material architectures with specific feature. Mesogenic materials possessing orientational order provide additional features to phase separated morphologies, such as bi-continuous or droplet. Here the matrix may be anisotropic and its interaction with isotropic drops may result in novel architectures such as colloidal crystals. Topological defects in the nematic matrix may positional order polymer drops in a perfect lattice. This self selected process emerges at specific interaction level between the mesogen and the polymer, whose value can be tuned by surfactants. Thus colloidal crystal formation in polymer/nematic mixtures is a sef-organizing material architecture unique to anisotropic soft matter.

Topological defects are an integral part of mesophases and arise due to frustration under non-planar confinement and under strong interaction between the phase separated sustrate and the mesogen. In the presence of binary nematic states, the role of non-mesogenic component in stabilizing the highly energetic defect core is significant. In this review we present recent thermodynamic models and calculation that describe the stability, composition, and geometry of defect cores by diffusion of non-mesogens. Results of this kind may be used to concentrate specific molecules or nanopartciles along defect lines, inside defect points, or at nematic-isotropic interfaces.

Blending colloidal particles into thermotorpic mesogens has been an active area in liquid crystal physics, and the effort has produced a significant number of material architectures and material systems. Currently another effort of producing nematic nanocomposites based on gold and other metallic nanoparticles and as well as carbon nanotubes is emerging.

In this review we have considered several approaches to thermodynamic modelling, from continuous solution thermodynamics, to discrete particle methods, suggesting that both approaches reveal complementary descriptions. Since facetted particles are ubiquitous, the review highlights new defect-particle superstructure, including string assemblies. The solution thermodynamic approach allows for positional order of the particles, and this fact is further analyzed in the discrete particle-mesogen model using triangular, square, pentagonal, and hexagonal geometries, revealing the presence of odd-even effects in particle arrangements due to different defect- particle interactions.

In summary, meso and macro scale thermodynamic modelling on mesogens mixtures with other mesogens, polymers, networks, solvents, colloidal and nanoparticles, provides a quantitative tool to develop new materials and devices. Mesogenic order enriches the structure o phase diagrams, the number and nature of mestabilities, the growth laws, and materials architectures. Future opportunities include multiscale multitransport, multidimensional modelling, new mesophases (chromonics, bend-core, dendrimer) and nanoparticle with well-designed ligand chemistries.

**Acknoeldegements**

This work is based on the research performed by A.D. Rey supported by the U.S. Office of Basic Energy Sciences, Department of Energy, grant DE-SC0001412. Le Fonds de recherche du Quebec - Nature et technologies and the National Research Council of Argentina supported E. R. Soule.